\documentclass[letter,apj]{emulateapj-rtx4}
\usepackage{amsmath,amssymb}
\pdfoutput=1
\usepackage{mathrsfs}
\usepackage{ulem}
\usepackage{hyperref}
\usepackage{mathtools}
\usepackage{graphicx}
\usepackage{natbib}
\usepackage{color}
\usepackage{bm}
\newcommand{\tokyo}{1}
\newcommand{\ias}{2}
\newcommand{\sagan}{3}

\altaffiltext{\tokyo}{University of Tokyo, Tokyo, Japan}
\altaffiltext{\ias}{Institute for Advanced Study, Princeton, NJ, USA}
\altaffiltext{\sagan}{NASA Sagan Fellow}

\begin{document}

\title{The Exoplanet Simple Orbit Fitting Toolbox (ExoSOFT):\\An open-source tool for efficient fitting of astrometric and radial velocity data}

\author{Kyle Mede\altaffilmark{\tokyo} and Timothy D.~Brandt\altaffilmark{\ias,\sagan}  }

\email{kylemede@astron.s.u-tokyo.ac.jp}

\begin{abstract}
We present the Exoplanet Simple Orbit Fitting Toolbox (ExoSOFT), a new, open-source suite to fit the orbital elements of planetary or stellar mass companions to any combination of radial velocity and astrometric data.  
To explore the parameter space of Keplerian models, ExoSOFT may be operated with its own multi-stage sampling approach, or interfaced with third-party tools such as emcee.  
In addition, ExoSOFT is packaged with a collection of post-processing tools to analyze and summarize the results.
Although only a few systems have been observed with both the radial velocity and direct imaging techniques, this number will increase thanks to upcoming spacecraft and ground based surveys.  Providing both forms of data enables simultaneous fitting that can help break degeneracies in the orbital elements that arise when only one data type is available.  The dynamical mass estimates this approach can produce are important when investigating the formation mechanisms and subsequent evolution of substellar companions.  ExoSOFT was verified through fitting to artificial data and was implemented using the Python and Cython programming languages; available for public download at https://github.com/kylemede/ExoSOFT under the GNU General Public License v3.
\end{abstract}


\section{Introduction}

Substellar companions exist in a wide array of masses and orbital configurations.  Confirmed companions\footnote{http://exoplanet.eu}\footnote{http://exoplanets.org/}\footnote{http://exoplanetarchive.ipac.caltech.edu/} have orbital separations ranging from 0.0044AU to 2500AU (for PSR 179-14 \citep{bailes} and WD 0806-661B b \citep{rodrigues} respectively), and masses as low as 0.022M$_{\Earth}$ (PSR B1257+12 \citep{Wolszczan538}).  This broad spectrum has highlighted the somewhat arbitrary nature of mass-based categorizations.  The canonical definition of a star is an object with sufficient core pressures for stable hydrogen burning, or a minimum mass of $\sim73-89$M$_{\rm J}$ depending on metallicity \citep{chabrier2000,grossman1974}. 
Bodies sufficiently massive to fuse deuterium, but not hydrogen, with masses from $\sim$13-80M$_{\rm J}$, are referred to as brown dwarfs; objects of even lower mass are called planets.  Deuterium burning has a negligible effect on a brown dwarf's evolution; a more physically interesting distinction between brown dwarfs and planets might be their mechanism of formation \citep{chabrier2014}.

Formation mechanisms are thought to be broken up into two general regimes: large objects form like stars, from the `top down', while planets grow from the `bottom up'.  Stellar mass companions are believed to form early on through the cooling and contraction of a fragment of the primary's protostellar cloud, or slightly later from a fragment of its protoplanetary disk  \citep{kuiper1951,cameron1978,boss1997,boley2009,vorobyov2013}.  Giant planets are thought to form when a rocky planet becomes sufficiently massive to accrete gas directly from the circumstellar disk \citep{safronov1969,goldreich1973,mizuno1980,pollack1996}. 
Direct collapse and core accretion operate over different, though likely overlapping, mass and separation regimes \citep{mordasini2012,Bate_2003,Bate_2009}; they may also produce different distributions of eccentricity.  Observational studies at these overlapping mass regimes would benefit from the precise and accurate planet masses obtained by fitting their dynamics.

Planets exist with a wide range of properties and their current orbits hold clues to their formation and evolutionary past.  Unfortunately, no single observing technique can provide data sufficient to fully model a companion's orbit and constrain all six of its orbital elements.  Two of those commonly applied for detecting substellar companions are the radial velocity method and high-contrast direct imaging.  Partial solutions to the orbit can be achieved by fitting the measured radial velocity or astrometry.
While the full set of orbital elements may be obtained by fitting both simultaneously for a single object.
Solutions found in this manner are capable of measuring the companion's dynamical mass, rather than the m$_{2}\sin(i)$ value estimated with the RV method alone.  Thus, the $\sin(i)$ degeneracy may be broken, enabling separate constraints on the inclination and companion mass.  However, no general open-source tools to perform joint RV and astrometric data fitting currently exist. 

Of the nearly 2000 confirmed exoplanets\footnote{http://exoplanetarchive.ipac.caltech.edu/}, $\sim$600 were found through radial velocity measurements and $\sim$40 with direct imaging, although none have data of both forms.  While a small number of objects with simultaneous coverage and estimated dynamical masses exist, they are all in the low-mass star and high mass brown dwarf range \citep[i.e.][also: J. Hagelberg et al. 2016, in preparation]{cre12a,cre12b,cre13a,cre13b,cre14,ryu16}.  Given the uncertainties in the atmospheric models of exoplanets and low-mass brown dwarfs, and the variety of orbital evolution scenarios, accurate dynamical mass values are important.  The advancements in ground- and space-based instrumentation will further populate the lists of known low mass companions observed with both techniques.  With Gaia precisely measuring acceleration in the plane of the sky for $\sim$10$^9$ stars, it is expected to lead to the discovery of thousands of planets astrometrically.  In addition, high-dispersion spectrographs such as IRD \citep{IRD}, CRIRES \citep{CRIRES} and SPIRou \citep{artigau2011}, will measure the radial velocity of low-mass stars with poor or no orbital coverage from previous instruments.  For those systems with separations and predicted contrasts detectable with next generation integral field spectrographs, including CHARIS \citep{peters2012}, SPHERE \citep{claudi2006}, and GPI \citep{macintosh2014}, follow-up observations can help investigate their atmospheres.  The higher sensitivities of all these instruments and spacecraft will increase the number of known companions and expand the overlap range of those detectable in both the radial velocity and astrometric data.

In this paper we introduce a new, open-source software package, ExoSOFT, for performing joint fitting of a system's orbital elements.  We have organized the paper as follows.  Section \ref{sec:models} covers the joint 3-dimensional model used, the background on fitting models of this kind and the approach ExoSOFT utilizes in Section \ref{sec:fitting}, and the implementation and verification of its capabilities for a synthetic example system in Section \ref{sec:implementation}.

ExoSOFT is available for public download from GitHub\footnote{https://github.com/kylemede/ExoSOFT} under the GNU General Public License v3, and has also been placed on the Python Package Index\footnote{https://pypi.python.org/pypi/ExoSOFT} for simple installation. 
{\bf The release version of the codebase used to produced the results in Section \ref{sec:verification} has been archived on Zenodo\footnote{http://doi.org/10.5281/zenodo.259500}.}

\section{Modeling the Orbit}\label{sec:models}

The classical solution to the two-body problem in Newtonian gravity is given by using Kepler's laws; six parameters are needed to fully describe the orbits (eg. \citet{heintz}).  We implement this approach adding some intermediate parameters to facilitate the calculation and comparison to measured radial velocity and astrometry.  Our full set of parameters is given in the top section of Table \ref{tbl:modelPars}; these are used to calculate the additional parameters in its lower section. 

\begin{deluxetable}{cc}  
\tablewidth{0pt}
\tablecaption{Model Parameters\label{tbl:modelPars}}
\tablehead{
    \colhead{Symbol} & 
    \colhead{Parameter}}
\startdata
$\varpi$ & Parallax \\
$P$ & Orbital period\\
$m_1$ & Mass of the primary star\\
$m_2$ & Mass of the companion\\
$\Omega$ & Longitude of the Ascending Node\\
$\omega$ & Argument of Periapsis\\
$i$ & Inclination\\
$e$ & Eccentricity\\
$T_o$ & Time of Last Periapsis\\
$\gamma$ & Instrument velocity offset \\
\hline
$m_{\rm tot}$ & Total mass ($m_1+m_2$)\\
$a_{\rm tot}$ & Total semi-major axis ($a_{1}+a_{2}$)\\
$\theta$ & True Anomaly\\
$E$ & Eccentric Anomaly\\
$M$ & Mean Anomaly\\
$K$ & Radial Velocity Semi Amplitude
\enddata
\end{deluxetable}

The predicted radial velocity and astrometry values, to be compared to the measured data for epoch $t$, are calculated in the following way.  The equations for the projected line-of-sight velocity, or radial velocity $v$, and projected locations in the plane of the sky, the astrometry, both require the anomalies $M$, $E$ and $\theta$.  Thus, the anomalies are determined first, followed by $v$, ending with the Thiele-Innes equations to get relative right ascension, $\Delta \alpha$ and declination, $\Delta \delta$.

The Eccentric Anomaly $E$ and True Anomaly $\theta$ are found starting from the Mean Anomaly $M$
\begin{equation}
M = \frac{2\pi}{P}\left(t-T_o \right)
\end{equation}
and its relation to $E$ through Kepler's equation is
\begin{equation}\label{eq:keplers}
M=E-e\times \sin(E),
\end{equation}
where $e$ is the eccentricity.
We solve Equation \eqref{eq:keplers} with Newton's method and use the resulting value to calculate $\theta$ following
\begin{equation}
\theta^\prime = \cos^{-1}\left(\frac{\cos(E)-e}{1-e\times \cos(E)}\right)
\end{equation}
and
\begin{equation}\label{eq:thetaFix}
    \theta = \left\{ \begin{array}{l l} \theta^\prime & E\leq \pi \\ 2\pi-\theta^\prime & E>\pi \end{array} \right. .
\end{equation}

Equations \eqref{eq:keplers}-\eqref{eq:thetaFix} provide the two key time dependent values, $E$ and $\theta$.

The radial velocity as measured for the primary due to the companion's motion is
\begin{equation}\label{eq:rvStandard}
v(t) =  K_1\left[\cos(\theta(t)+\omega_1)+e \cos(\omega_1)\right]+\gamma
\end{equation}
where $\omega_1$ is the Argument of Periapsis for the primary, $\gamma$ is the instrument velocity offset and the semi amplitude of the primary ($K_1$) is 
\begin{equation}\label{eq:K}
K_1 = \left[\frac{2\pi G}{P}\right]^{1/3}\frac{m_2\sin(i)}{\left(m_{\rm tot}\right)^{2/3}}\frac{1}{\sqrt{1-e^2}} 
\end{equation}
with $m_2$ being the companion's mass, $m_{\rm tot}$ the total mass of both objects and $P$ the period.  When only radial velocity data are available, Equation \eqref{eq:K} shows that the companion's mass and inclination cannot be independently constrained.

The Thiele-Innes method is a standard approach to compute the predicted astrometric values given a set of input orbital elements \citep{Thiele,Van}, using a set of intermediate quantities

\begin{align}
\label{eq:THa}
A &= a_{\rm tot}\left[\cos(\Omega_2)\cos(\omega_2)-\sin(\Omega_2)\sin(\omega_2)\cos(i)\right]\\
\label{eq:THb}
B &= a_{\rm tot}\left[\sin(\Omega_2)\cos(\omega_2)+\cos(\Omega_2)\sin(\omega_2)\cos(i)\right]\\
\label{eq:THf}
F &= a_{\rm tot}\left[-\cos(\Omega_2)\sin(\omega_2)-\sin(\Omega_2)\cos(\omega_2)\cos(i)\right]\\
\label{eq:THg}
G &= a_{\rm tot}\left[-\sin(\Omega_2)\sin(\omega_2)+\cos(\Omega_2)\cos(\omega_2)\cos(i)\right] .
\end{align}
These equations are in the form that is standard in the field \citep{heintz,aitken}.
In Equations \eqref{eq:THa}-\eqref{eq:THg}, the total semi-major axis of the apparent ellipse ($a_{\rm tot}$) is calculated with Kepler's third law,
\begin{equation}\label{eq:K3}
a_{\rm tot} = \left[\frac{P^2G m_{\rm tot}}{4\pi^2} \right]^{1/3}
\end{equation}
with $G$ in Equation \eqref{eq:K3} being Newton's gravitational constant.

Equations \eqref{eq:THa}-\eqref{eq:THg} give the relative location of the secondary with respect to the primary in the plane of the sky.  The matching relative right ascension $\Delta \alpha$ and declination $\Delta \delta$ for comparison to the data are found with
\begin{equation}
\label{eq:th-I-Dec}
\Delta \delta = AX(t)+FY(t)
\end{equation}
\begin{equation}
\label{eq:th-I-RA}
\Delta \alpha = BX(t) + GY(t)
\end{equation}
given
\begin{equation}
\label{eq:28-1.5a}
X(t) = \cos(E(t))-e
\end{equation}
\begin{equation}
\label{eq:28-1.5b}
Y(t) = \sqrt{1-e^2}\sin(E(t)) .
\end{equation}

The orbits of the two bodies $m_1$ and $m_2$ have matching values for all the orbital elements except $\Omega$ and $\omega$.  To convert between them, the relations $\omega_1 = \omega_2 +\pi$ and $\Omega_1 = \Omega_2 +\pi$ are used.  For all cases, we assume the measured astrometry to be the relative positions of the two bodies tracing out an apparent ellipse on the sky, which reduces to the orbit of the companion when $m_1 \gg m_2$.

The equations in this section calculate a set of observables ($v$, $\Delta \alpha$, and $\Delta \delta$) that we then compare to the corresponding measurements.

\section{Model Fitting}\label{sec:fitting}
\subsection{Method}\label{sec:basicTheory}

Model fitting can be broken down into two main tasks: solving for a set of parameters approaching those with the maximum likelihood, and estimating their uncertainties.  In a Bayesian framework the full set of information is contained within the posterior probability distribution of the model parameters, $p({\rm Model}\vert {\rm Data})$.  \citet{ford2005} \& \citet{ford2006} discuss the use of Bayesian inference to solve for this distribution in the case of fitting observations of exoplanets to Keplerian models; for details beyond those given here, we refer the reader to those papers.  

The posterior probability distribution, $p({\rm Model}\vert {\rm Data})$, is computed using Bayes' theorem, 
\begin{equation}
p({\rm Model}\vert {\rm Data}) \propto p({\rm Data}\vert {\rm Model}) p({\rm Model}).
\end{equation}
Here $p({\rm Data}\vert {\rm Model})$ is the likelihood of the model parameters, $\mathscr{L}({\rm Model})$, and $p({\rm Model})$ is the prior probability of the parameters based on previous knowledge.  Assuming that the errors in the observed data are independent and Gaussian distributed, the likelihood can be expressed in terms of the usual $\chi^2$:
\begin{equation}\label{eq:likelihood}
	\mathscr{L}\left({\rm Model}\right) = p({\rm Data} \vert {\rm Model}) \propto \exp\left(-\frac{\chi^2}{2}\right)
\end{equation}
with $\chi^2$ being a sum over data points $i$,
\begin{equation}\label{eq:chiSquared}
 \chi^2 = \sum_{i}\frac{\left({\rm Model}_i-{\rm Data}_i\right)^2}{\sigma^2_i}
\end{equation}

The posterior probability distribution can then be computed by finding the value of $\mathscr{L}({\rm Model})p({\rm Model})$ for all possible parameter combinations and integrating.  Due to the high dimensionality of Keplerian models though, the parameter space is large, making this brute-force approach computationally problematic.  A variety of routines exist for sufficiently sampling the posterior without calculating its value everywhere.  In Section \ref{sec:MCMC} we will discuss the approaches used in ExoSOFT, particularly that of Markov Chain Monte Carlo, along with the specific priors for each of the model parameters.

\subsection{MCMC}\label{sec:MCMC}

Forming the posterior distributions can be achieved by integrating Bayes' theorem over the model's parameter space.  Depending on the form of the likelihood function and the dimensionality of the model space, evaluating this integral directly or through `shotgun' Monte Carlo can prove computationally impractical.  One alternative is to draw statistically dependent samples with Markov Chain Monte Carlo (MCMC) \citep{liu}.  A Markov Chain is a sequence of points satisfying the dependency defined by the Markov property: each point in the chain is a function only of its immediate predecessor with no memory of the chain history.  MCMC guarantees convergence to the posterior probability distributions in the limit of a large number of steps.  A widely used algorithm to form such chains is that of Metropolis-Hastings, originally detailed in \citet{metropolis} and further generalized by \citet{hastings}.  Unfortunately, the convergence rate is typically impossible to compute; we discuss convergence in more detail in Section \ref{sec:MCMCcaveats}.

There are multiple approaches to sampling the posterior distributions of the Keplerian model in Section \ref{sec:models} with ExoSOFT.  One is to use a built-in MCMC sampler based on the Metropolis-Hastings algorithm.  This implements the standard form of the rejection function,
\begin{equation}\label{eq:Ma3}
	r\left(\bm{\xi},\bm{\xi}_{j+1}\right) = min \left\{1, \frac{\mathscr{L}\left(\bm{\xi}_{j+1}\right)}{ \mathscr{L}(\bm{\xi}_j)} \frac{p\left(\bm{\xi}_{j+1}\right)}{p\left(\bm{\xi}_j\right)}    \right\},
\end{equation}
where $\bm{\xi}_{j+1}$ is a step, a set of parameters; with this, the likelihood is $\mathscr{L} = \mathscr{L}\left(\rm Model_{j}\right) = \mathscr{L}\left(\bm{\xi}_j\right)$, and the prior probability is $p = p\left(\rm Model \right) = p\left(\bm{\xi}_j \right)$. 
The proposed steps are drawn from normal distributions of fixed width in each parameter centered on the current step.  Following the Gaussian error assumption in Section \ref{sec:basicTheory}, the likelihood ratio is
\begin{equation}\label{eq:likelihoods}
    \frac{\mathscr{L}\left(\bm{\xi}_{j+1}\right)}{\mathscr{L}\left(\bm{\xi}_j\right)} = \exp\left(-\frac{{\chi}^{2}_{\bm{\xi}_{j+1}} - {\chi}^{2}_{\bm{\xi}_{j}} }{2}\right) .
\end{equation}
ExoSOFT also provides $\mathscr{L}$ and $p$ in a format that easily interfaces with MCMC packages like emcee \citep{emcee}, using the combined function $ln\left(\mathscr{L}\left(\bm{\xi}\right)\right) + ln\left(p\left(\bm{\xi}\right)\right)$.
As such, emcee has been made a dependency of ExoSOFT and can be selected for use with a flag in the settings file.

We chose prior probabilities, or `priors', for each parameter with simplicity in mind.  These should only be considered as suggestions, and may be easily changed in the ExoSOFT settings files to suite the user's preferences or to take into account the results of more recent work.  The complete set of parameters in our full joint analysis are ($m_1$,$m_2$,$\varpi$,$P$,$i$,$e$,$T_o$,$\Omega$,$\omega$,$\gamma$).  Independent of their distribution, the priors are all given wide boundary limits, each available for change within the settings.  Those of $\Omega$ and $\omega$ are drawn without limits, but the stored values are then shifted into [0,2$\pi$].

For $m_1$, the prior can be represented by the Initial Mass Function (IMF) or Present Day Mass Function (PDMF).  \citet{chabrier2003} found functional forms of these for a range of stellar populations, including those of the galactic disk.  For $m_2$ however, particularly for stellar binaries, the IMF may be a poor approximation to the companion mass function (CMF); that of \citet{metchev2009} might be more appropriate.  More recent investigations \citep[e.g.][]{brandt2014} were consistent with \citet{metchev2009} down to a few Jupiter masses.
PDMF and IMF can be found in Table 1 of \citet{chabrier2003}, and that of the CMF in Equation (8) of \citet{metchev2009}.  By default we use the PDMF for the primary and CMF for the companion, with the IMF being an option also available to the user for either body.

The prior for the parallax ($\varpi$) may be derived from the assumption that stars are uniformly distributed in space.  For nearby objects, within a few 100 pc, this prior is
\begin{equation}\label{eq:paraPrior}
	p\left(\varpi\right) \propto \frac{1}{\varpi^4} .
\end{equation}
In cases where high S/N distance measurements are available, this information can be included with a Gaussian prior with center and width equal to its distance and error, respectively.  These two components would then be multiplied to form a combined prior for the parallax. For distant objects, over a few hundred pc, the assumption of uniform stellar distribution is invalid.  Low signal-to-noise measurements of the parallax, in combination with Equation \eqref{eq:paraPrior}, would lead to unrealistically large distances.  ExoSOFT currently requires accurately measured parallaxes.  In practice, these are available for nearby bright stars from {\it Hipparcos} \citep{vanLeeuwen_2007}, and the sample of objects with well-measured parallaxes will soon expand dramatically thanks to Gaia.

The adopted prior for the period is a power-law, $p\left(P\right) \propto P^{\gamma}$.  Radial velocity surveys favor $\gamma \approx -0.7$ \citep{cumming2008}, while $\gamma = -1$ gives equal numbers per logarithmic period.

We assume systems to be randomly oriented with $p\left(i\right) \propto \sin(i)$.

Following the results of investigations into the eccentricity distributions of low-mass companions \citep[e.g.][]{wang2011, shen2008,kipping2013,kipping2016,hogg2010,shabram2016,duquennoy1991,juric2008,cumming2008}, ExoSOFT offers multiple priors for the eccentricity.  Currently, these include: uniform, Rayleigh, Rayleigh+Exponential, Beta, $p(e)=2e$ for companions with ($P>$1000 days), and that of \citet{shen2008}.  Additional priors can be added and modified manually by the user.  
This can be accomplished by placing a customized version of the priors module in the user's current working directory.  

The remainder of the parameters, ($T_0$, $\Omega$, $\omega$, $\gamma$), have been given uniform priors.

\subsection{Tuning MCMC}\label{sec:MCMCcaveats}

Given an infinitely long Markov chain, convergence to the posterior probability distributions is guaranteed.  In practice when trying to integrate Bayes' theorem for a particular model, one of the primary goals is to minimize the number of samples necessary to achieve sufficient convergence.  Therefore, before performing MCMC, a metric to measure its convergence is required along with steps to optimize the convergence rate.

It is difficult to determine if a Markov Chain has converged to the posterior probability distribution.  From a given set of parameter values, there is no guarantee that the likelihood function will smoothly increase towards the global maximum of the posterior.  In cases with complicated likelihood topography, there might be many places where a chain could become stuck.  Chains started near a local maximum could falsely appear to have converged or take a long time to diffuse into other regions.  For this reason, the specific starting position of a Markov chain, and any following samples drawn far from the likelihood peaks, must be removed to avoid any disproportionalities in the posterior distributions they might induce.  This initial period is referred to as the `burn-in'.  While the exact number of steps involved is not clearly defined, \citet{tegmark} took it to be over when the likelihood in a particular chain was equal to the median likelihood of all chains combined.

The rate at which a Markov chain explores the posteriors is related to the width, or $\sigma$, of each parameter's proposal distribution 
.  For large $\sigma$, proposed steps tend to lie far from the current one and in a region where they are unlikely to be accepted.  By contrast, a small $\sigma$ would lead to a high rate of acceptance, but a slow rate at which the posteriors are explored.  \citet{Gelman} showed that the highest diffusion speed of a one dimensional model was achieved when the probability of accepting a proposed step was 0.441.  They tested a variety of multi-dimensional models concluding that an acceptance rate of $\sim$25\% is suitable for higher dimensional models and $\sim$50\% for those with only 1 or 2 dimensions.   

With a Markov sampling scheme the dependence between neighboring samples can dramatically reduce the number of effectively independent steps taken.  Unfortunately, prior to running a Markov chain, there are no direct methods to predict how many independent steps will be taken over a certain number of samples.  Instead, following the completion of all Markov chains, a common practice is to estimate if enough steps were taken using either the Gelman-Rubin statistic \citep{gelman1992}, 
or the integrated autocorrelation time \citep{goodman}.  As discussed in \citet{emcee} and \citet{goodman}, the integrated autocorrelation time can be represented by
\begin{equation}\label{eq:autocorr}
\tau = 1+ 2 \sum^{\infty}_{T=1} \frac{C\left( T \right)}{C\left(0 \right)} ,
\end{equation}
where $T$ is the time lag necessary for covariance between samples, $C(T)$, to go to zero, ensuring their independence.  With this, the number of effectively independent steps in a Markov chain of length $L_c$  is $\sim L_c/\tau$.

\section{Implementation and Verification}\label{sec:implementation}

In this section we describe how the model discussed in Section \ref{sec:models} and types of parameter space exploration in Section \ref{sec:fitting} are implemented.  The bulk of the code was written in the Python programming language, with the more computationally expensive model in Cython.  Implementing the model in this way makes it run 5-30 times faster than the same code written purely in Python.  ExoSOFT can be found for public download at https://github.com/kylemede/ExoSOFT.

\subsection{Simulation Stages}

ExoSOFT offers four of its own algorithms for exploring the parameter space, each for a different task: standard `shotgun' Monte Carlo, Simulated Annealing, Sigma Tuning and MCMC.  It is also ready to be interfaced with third-party sampling tools.

As discussed in Section \ref{sec:MCMC}, standard Monte Carlo is suitable for models with few free parameters, or when the likelihood function is not strongly peaked.  In such cases, it could in fact sample the posterior probability distributions significantly faster than MCMC.  With each sample being completely independent, it can also be useful for getting an idea of where the regions of minimum $\chi^2$ lie. 

In cases where standard Monte Carlo is inefficient, Markov sampling techniques are suggested.  To minimize the burn-in, suitable starting positions for the Markov chains are found using Simulated Annealing.  As discussed in \citet{SimAnneal}, Simulated Annealing also makes use of the Metropolis-Hastings algorithm with an additional `temperature' factor in the likelihoods ratio, modifying this ratio in the rejection function to
\begin{equation}\label{eq:beta}
	\left( \frac{\mathscr{L}\left(\bm{\xi}_{j+1}\right)}{\mathscr{L}\left(\bm{\xi}_j\right)} \right)^{1/T} = \exp \left( -\frac{{\chi}^{2}_{\bm{\xi}_{j+1}} - {\chi}^{2}_{\bm{\xi}_{j}}}{2T} \right)  .
\end{equation}  
At high temperatures Equation \eqref{eq:beta} is temperature dominated, and likely to accept even unfavorable steps, thereby mitigating any local topographical bumps encountered.  A cooling scheme started at a sufficiently high temperature ($\gtrapprox 100$) explores a large portion of the parameter space and is slowly `annealed' 
towards the parameter set with the global maximum likelihood.  In our implementation, the chains are repeatedly heated and cooled, each time from a lower starting temperature, while maintaining a list of the samples with the highest likelihoods
achieved so far.  We stop this process when the fits have all converged to a single set of parameter peaks in the posterior probability distributions.

After finding an initial set of parameters, ExoSOFT can sample the posterior probability distributions with either MCMC or the affine-invariant ensemble sampler built into emcee \citep{emcee}.  When using MCMC, we first apply a version of the Metropolis-Hastings algorithm to determine the appropriate widths for each proposal distribution.  
Periodically the acceptance rates are calculated, typically after $\sim$200 samples per varied parameter, and the fixed $\sigma$ values of the normal proposal distributions for each parameter incremented accordingly until stable acceptance rates between 25-35\% are reached; this process is occasionally referred to as Sigma Tuning. 
In contrast to the earlier Simulated Annealing and later MCMC stages, Sigma Tuning only requires a single chain and has been implemented in such a way that it automatically completes once the acceptance rates stabilize.  In cases where the observational data is sparse, or parameters highly correlated, the use of an ensemble sampler has been proven to increase the sampling efficiency \citep{emcee, goodman}.  
Thus, our recommended approach is to utilize
the starting position produced by Simulated Annealing to initialize the walkers of the affine-invariant ensemble sampler built into emcee \citep{emcee}.

The solutions for determining a set of initial parameters and tuning the proposal distribution widths lend themselves well to integration with other third-party sampling tools, including PyMC \citep{pymc} and PyMultiNest \citep{MULTINEST}.

\subsection{Verification}\label{sec:verification}

Our model was verified with synthetic data produced using an independent Keplerian tool built upon the PyAstronomy package\footnote{Found at https://github.com/sczesla/PyAstronomy}.  The observable values had their errors realized from Gaussian distributions, using a percentage of each observable's absolute mean as the distribution widths.  Such synthetic data were produced for various systems spanning the range of exoplanets and binary star systems observed to date to ensure ExoSOFT's ability to fit any system of interest.  

For demonstration purposes, the approximate orbital elements of Jupiter were used assuming a host star at a distance of 20 pc, or a parallax of 50 mas, with the orbital plane inclined by 45$^{\circ}$ to Earth's line-of-sight.  A variety of measurement errors were assumed for the same orbital elements to show the convergence of the resulting posterior distributions with improving data quality; the errors indicated are a percentage of each observable parameter's absolute mean value.

\begin{figure*}
\begin{center}
\includegraphics[scale=0.48,trim=0 0.5cm 0 0.4cm,clip]{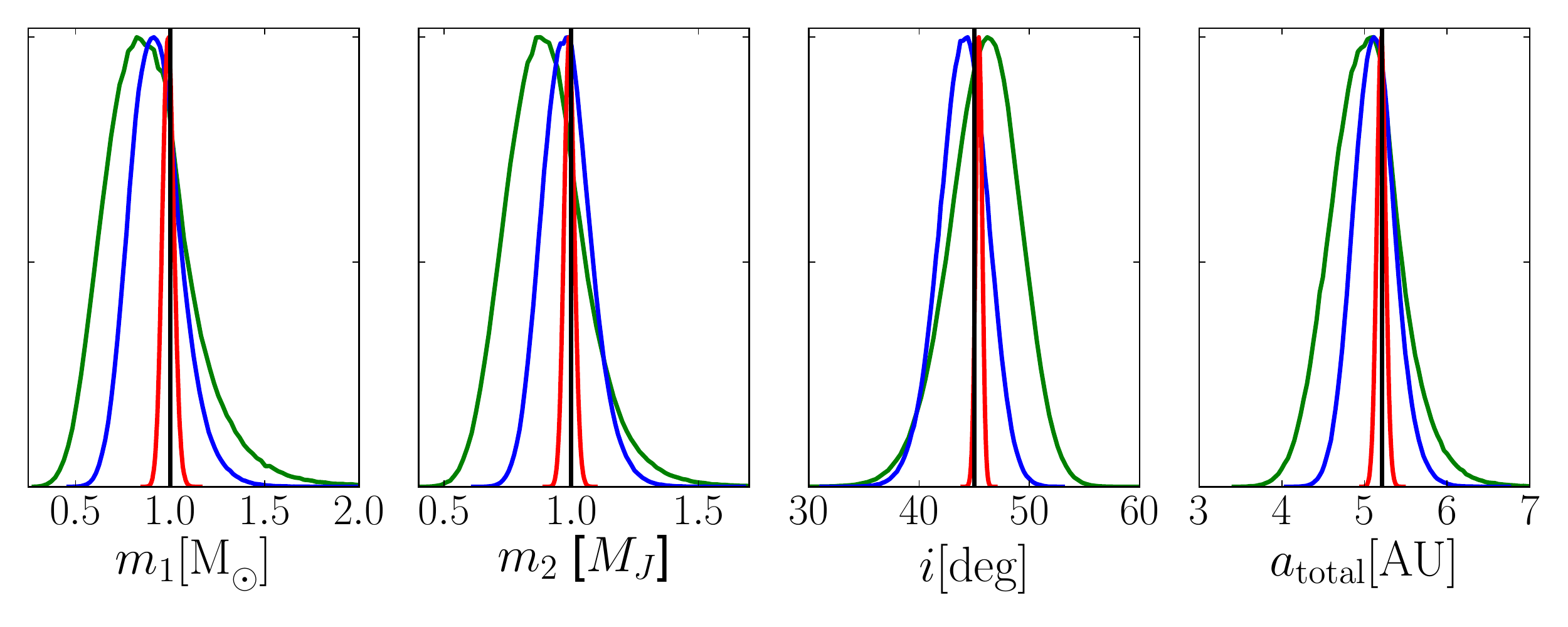}
\caption{ The convergence of ExoSOFT's marginalized posterior probability distributions to noisy synthetic data.  The observational errors in astrometry and radial velocity are assumed to be a percentage of the absolute mean value for each observable ($\Delta \delta$, $\Delta \alpha$, $v$).  We draw a realization of the noise for each measurement from a normalized Gaussian of given width.  The lines shown are for 10, 5 and 1\% errors, corresponding to the colors green, blue and red.  The assumed true values are indicated with black vertical lines.\label{fig:realizationsConvergence} }
\end{center}
\end{figure*}

\begin{deluxetable}{lcccr}
\tablewidth{0pt}
\tablecaption{Synthetic Jupiter Fit Results\label{tbl:jupiterTable}}
\tablehead{
    Parameter & 
    Synthetic & 
    Best-Sample&
    Median & 
    1$\sigma$ Range
}
\startdata
$m_1$ [$M_{\odot}$] & 1 & 0.87 & 0.93  &  0.13\\
$m_2$ [$M_{\rm J}$] & 1 & 0.942 & 0.985 &  0.094\\
$\varpi$ [mas] & 50 & 51.4 & 50.3 & 2.3\\
$\Omega_2$ [$^{\circ}$] & 100.6 & 101.4 & 101.7 &  1.1\\
$e$ & 0.048 & 0.051 & 0.042 &  0.014\\
T$_0$ [MJD]& 50639 & 50670 & 50660  &  140\\
$P$ [Yrs] & 11.9  & 11.94  & 12.05 &  0.18\\
$i$ [$^{\circ}$] & 45 & 43.7 &  44.2 &  2.2\\
$\omega_2$ [$^{\circ}$] & 14.8 & 17 &  16 &  12\\
$a_{\rm tot}$ [AU] & 5.21 & 4.98 & 5.13 & 0.25\\
$\gamma$ [m/s] &  0  & 0.03  & -0.06 &  0.16\\
$\chi^2$ ($\nu$) & 46.9 (35) 
\enddata
\end{deluxetable}

Figure \ref{fig:realizationsConvergence} shows the marginalized posteriors using realizations of 10\%, 5\% and 1\% Gaussian noise on the measured parallax, radial velocity, and angular separation.
The 68\% confidence regions shrink by a factor of $\sim$10 from the 10\% to 1\% cases, shown as the green and red lines in Figure \ref{fig:realizationsConvergence} respectively.
The posterior distributions of $T_0$ and $\omega$ (not shown) have larger widths and decrease faster with improving data quality; this is an artifact of our choice of a nearly circular orbit.
In all of the hypothetical Jupiter demonstration cases, ExoSOFT used a uniform prior for $e$, appropriate for planets.  This was implemented by drawing $\omega$ and $e$ simultaneously according to the parametrization from \citet{albrecht2012}.

\begin{figure*}
\begin{center}
\includegraphics[scale=0.5,trim=0 0.5cm 0 0.4cm,clip]{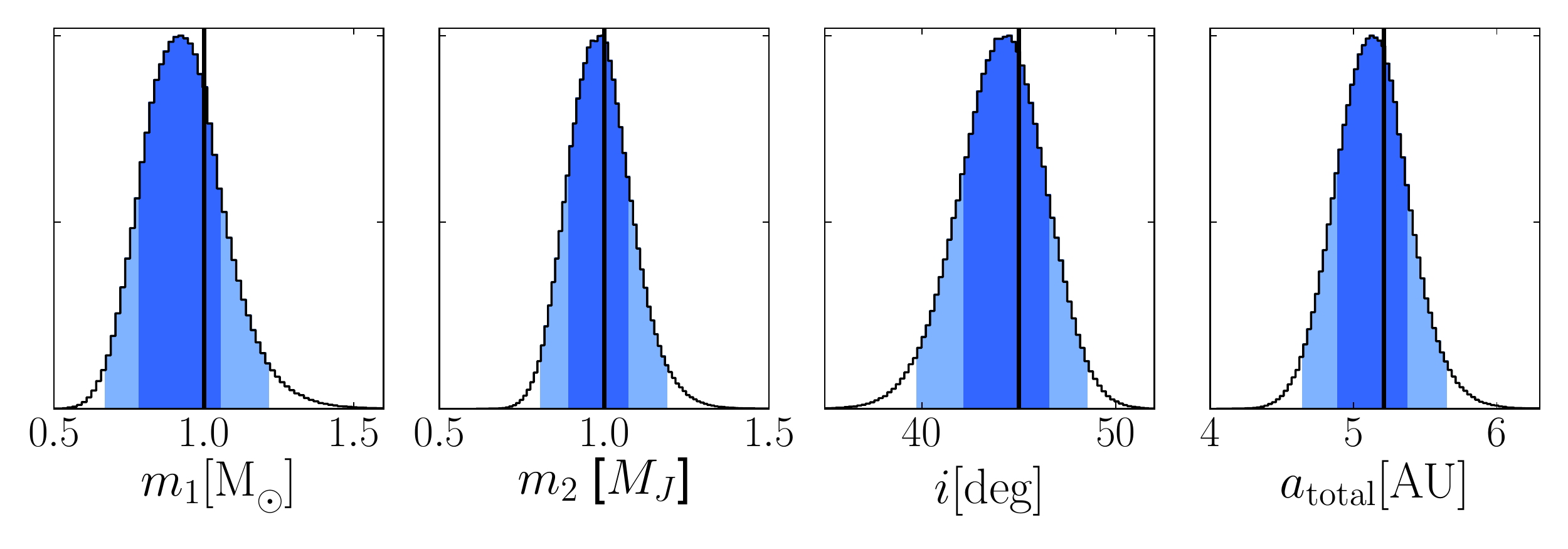}
\caption{ Marginalized posterior probability distributions for a subset of the parameters fit during verification with synthetic data of a Jupiter analogue with a 5\% realization of the errors in radial velocity and astrometry, at distance of 20 pc and inclination of $45^{\circ}$.  The darker blue represents the 68\% regions of confidence around the medians, with the lighter being that of 95\% confidence.  As expected, the vertical black lines representing the true values fall within our 68\% confidence regions.\label{fig:5percentPosteriors}}
\end{center}
\end{figure*}

\begin{figure*}
\begin{center}
\includegraphics[scale=0.65]{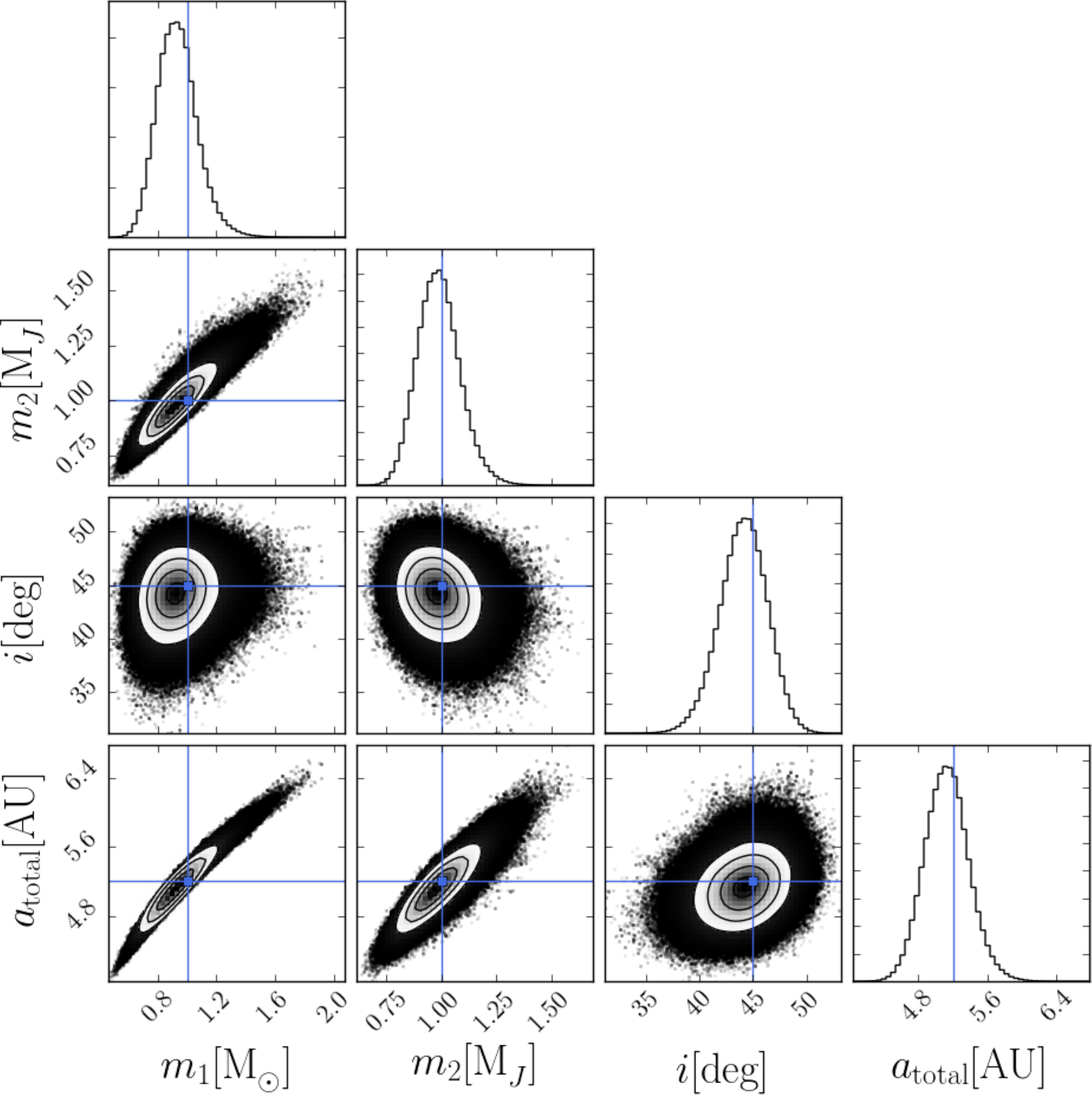}%0.666
\caption{ Corner plot for a subset of the parameters fit during verification with synthetic data of a Jupiter analogue with a 5\% realization of the errors in radial velocity and astrometry, at distance of 20 pc and inclination of $45^{\circ}$.  The solid blue lines represent the true values.  The covariance between the masses and total semi-major axis seen here arises from the direct use of Kepler's third law in our model discussed in Section \ref{sec:models}.\label{fig:corner}}
\end{center}
\end{figure*}

Figure \ref{fig:5percentPosteriors} shows the resulting marginalized posteriors for one realization of 5\% measurement errors; Table \ref{tbl:jupiterTable} provides a complete summary of the fit values.  The covariance between a subset of the orbital elements is represented in a corner plot in Figure \ref{fig:corner}.  Here the relationship between the masses and semi-major axis can be seen, arising from Kepler's third law in Equation \eqref{eq:K3} being directly included in the model equations.

\begin{figure*}
\begin{center}
  \includegraphics[width=0.4\textwidth]{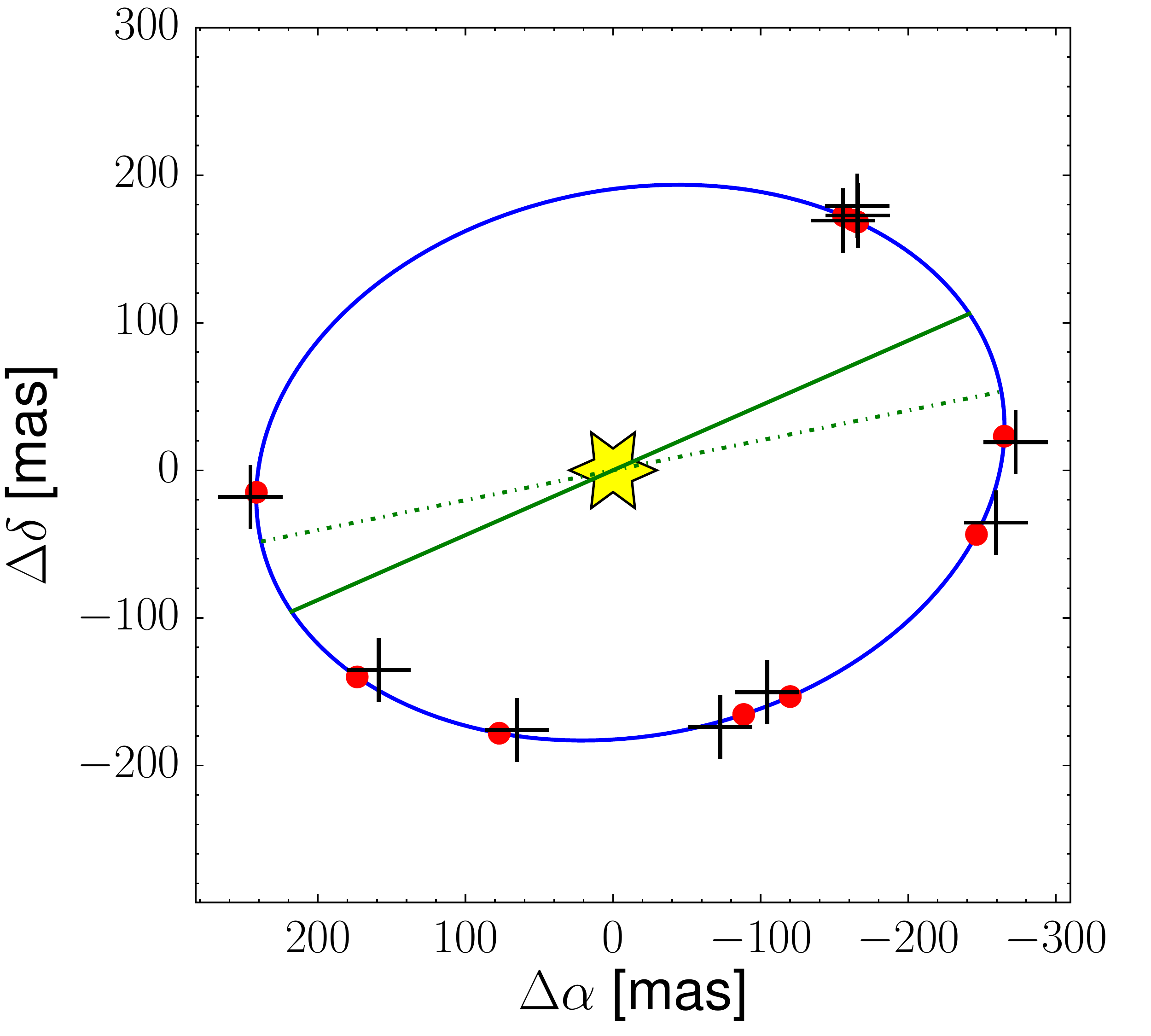}
  \includegraphics[width=0.59\textwidth]{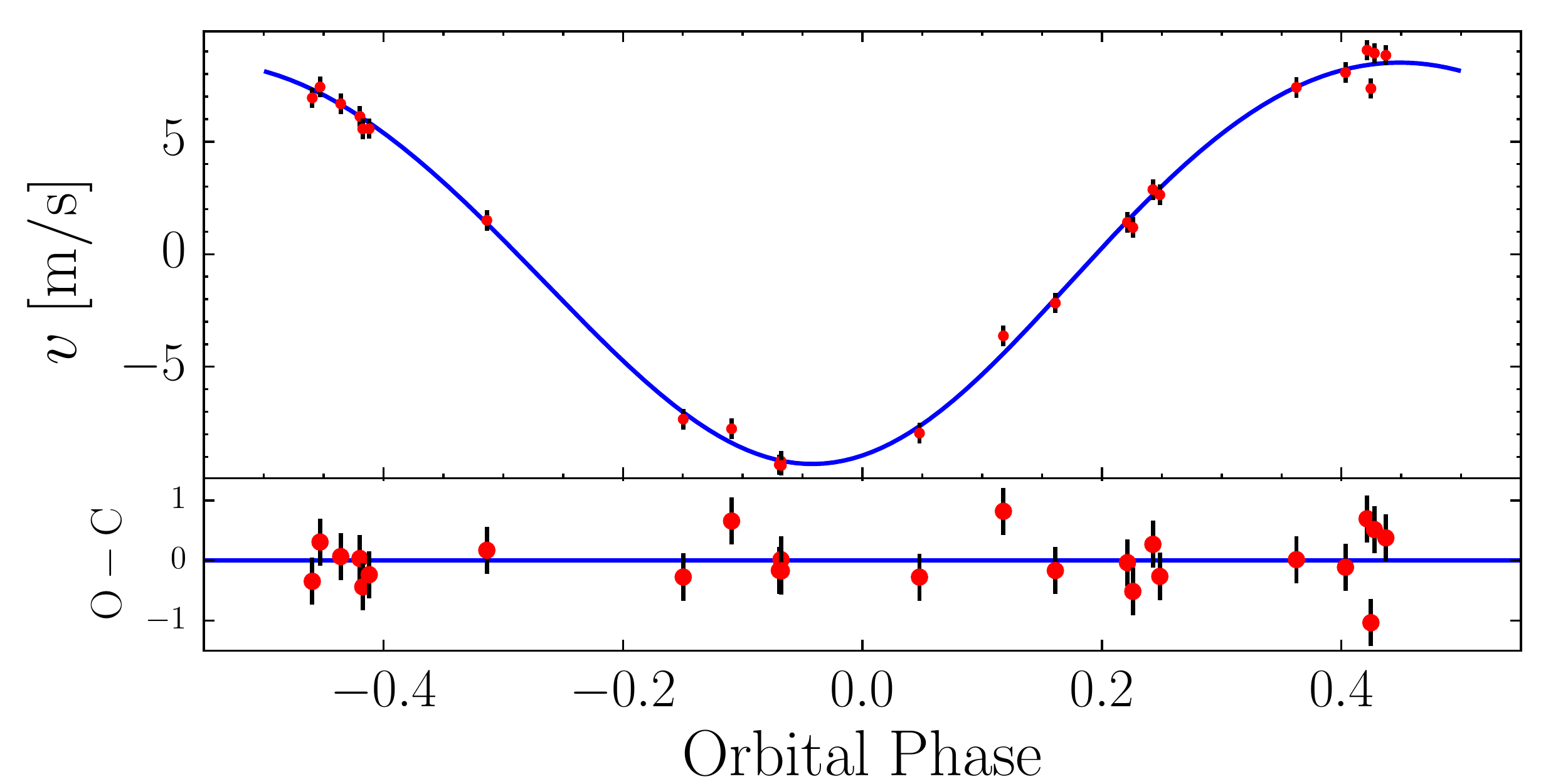}
\caption{ The orbit for the best sample
plotted atop the synthetic astrometry (left) and radial velocity data (right) of our Jupiter analogue test case with a 5\% realization of the errors.  In the astrometry fit the solid green line represents the projected semi-major axis and the dash-dot the line of nodes.  The red dots represent the predicted location and radial velocity values at the time of each observation.\label{fig:5percentOrbits}}
\end{center}
\end{figure*}

The best sample, defined as the set of stored parameters with the highest likelihood, has $\chi^2 = 46.9$ for 35 degrees of freedom (25 epochs of radial velocity data, 10 epochs of astrometry in RA and Dec, and 10 fitted parameters), for a reduced $\chi^2 = 1.34$.
Figure \ref{fig:5percentOrbits} shows the orbit for the best sample
plotted atop the noisy astrometry and radial velocity data.  The residuals, as expected, scatter about the best sample's orbit and typically lie $\sim$1$\sigma$ away.

The joint model given in Section \ref{sec:models} mutually constrains the parameters $P$, $m_{\rm tot}$, $\omega^*$, $i$, $e$ and $T_0$, where $\omega^*$ represents the value for either body following the relation $\omega_1=\omega_2+\pi$.  The other parameters ($\gamma$, $m_2$, $\Omega_2$, $\varpi$) are constrained by a single observation type, or in the form of a combined parameter, as is the case for $m_{\rm tot}=m_1+m_2$.
We performed the final fitting with the affine-invariant ensemble sampler of emcee and investigated the mixing and convergence of each parameter using the integrated autocorrelation time, discussed in Section \ref{sec:MCMCcaveats} and represented in Equation \eqref{eq:autocorr}.  The least convergent parameter in all three error realizations was $T_0$, with a total of $5 \times 10^6$ samples distributed over 1000 walkers, its $\tau$ was 690 for the 5\% case.
This slow convergence is an artifact of our choice of a nearly circular orbit (the definition of $T_0$ is arbitrary for a circular orbit) and is much less pronounced for more eccentric systems.

While this test case is at the edge of current instruments' sensitivity, the $\sim$0.$\!\!''2$ separation is within the expected limits of upcoming instruments like CHARIS \citep{CHARIS}, SPHERE \citep{claudi2006} and GPI \citep{macintosh2014}.  However, even with the improved contrast ratios expected at these separations, this hypothetical planetary system would need to be dramatically younger than Jupiter for detection through direct imaging techniques.  
The uncertainties in $m_1$ and $m_2$ also depend on the precision of the measured parallax, though this is already better than 1\% for many bright, nearby stars \citep{vanLeeuwen_2007} and will further improve with the forthcoming Gaia data release.

An example of ExoSOFT applied to real data can be found in \citet{V450And}.  In that paper we introduce the first direct imaging detections of the companion to V450 And, solve for its joint orbital solution and compare the resulting dynamical masses to those from stellar model fits based on the photometry.

\section{Conclusion}

We have presented ExoSOFT, a toolbox for the Keplerian orbital analysis of exoplanets and binary star systems.  It is free to compile, open source, fits any combination of astrometric and radial velocity data, and offers four of its own parameter space exploration techniques, along with an integration of the affine-invariant ensemble sampler in emcee \citep{emcee}.  Our suggestion is to first use Simulated Annealing to find a set of parameters near the global maximum likelihood.  This can then be followed by Sigma Tuning and MCMC to optimize and fill out the posterior probability distributions.  When data is sparse or parameters highly correlated, we instead recommend emcee to increase sampling efficiency. 
ExoSOFT is packaged together with a set of post-processing and plotting routines to summarize the results.  Verifications were performed by fitting noisy synthetic data produced with an independent Keplerian model for a variety of systems ranging the entire parameter space.  Examples of its use on real data include \citet{V450And}, for the V450~And system, and the $\tau$ Boo AB binary in \citet{Mede2014}.  

Simultaneous fitting to both radial velocity and astrometric data helps to break the degeneracies in Keplerian model parameters.  The dynamical mass estimates of the companion and the well-constrained values of the other orbital elements found in this way are important clues to substellar formation and evolution.  
The companion's mass and age may be used to place constraints on the input parameters of atmospheric models for brown dwarf and low mass stars.  The initial conditions of substellar objects remain uncertain, with theoretical predictions ranging from hot-start to cold-start scenarios \citep{marley2007}.
Mass, eccentricity and separation are also important to help distinguish between the variety of dynamical evolution possibilities in multi-planet systems.

Recent surveys aiming at the high-end range of substellar mass companions \citep{wilson2016} and at the lower end \citep{hebrard2016} show how the capabilities of new instruments are expanding the range of companions detectable with the radial velocity method.  Complementary to these, observations of HR 8799 with SPHERE and GPI \citep[eg.][]{zurlo2016,bonnefoy2016,apai2016,ingraham2014} highlight the improved astrometric and characterization capabilities with next generation integral field spectrographs.  Combining ground based facilities with the Gaia spacecraft will broaden not only the mass range, but also the number of objects simultaneously observed with both radial velocity and direct imaging techniques. 
With its capacity to simultaneously fit both of these data types, effectively explore the parameter space, and with its post-processing and plotting tools, ExoSOFT is well-suited to constrain the orbits of newly discovered systems.

\section{Acknowledgements}
K.M. gratefully acknowledges support from the Mitsubishi Corporation International Student Scholarship.  This work was performed in part under contract with the Jet Propulsion Laboratory (JPL) funded by NASA through the Sagan Fellowship Program executed by the NASA Exoplanet Science Institute.

This research made use of the SIMBAD literature database, operated at CDS, Strasbourge, France, and of NASA's Astrophysics Data System.  This research made use of the NASA/IPAC/NExScI Star and Exoplanet Database, which is operated by the Jet Propulsion Laboratory, California Institute of Technology, under contract with the National Aeronautics and Space Administration.  This research has made use of the Exoplanet Orbit Database and the Exoplanet Data Explorer at exoplanets.org.  This research has made use of the NASA Exoplanet Archive, which is operated by the California Institute of Technology, under contract with the National Aeronautics and Space Administration under the Exoplanet Exploration Program.

\pagebreak

\bibliographystyle{apj_eprint}
\bibliography{citations}

\begin{thebibliography}{}

\bibitem[\protect\citeauthoryear{{Aitken}}{{Aitken}}{1935}]{aitken}
{Aitken}, R.~G. 1935, {The binary stars}

\bibitem[\protect\citeauthoryear{{Albrecht} et~al.}{{Albrecht}
  et~al.}{2012}]{albrecht2012}
{Albrecht}, S., {Winn}, J.~N., {Butler}, R.~P., et~al. 2012, \apj, 744, 189

\bibitem[\protect\citeauthoryear{{Apai} et~al.}{{Apai} et~al.}{2016}]{apai2016}
{Apai}, D., {Kasper}, M., {Skemer}, A., et~al. 2016, \apj, 820, 40

\bibitem[\protect\citeauthoryear{{Artigau}, {Donati}, \& {Delfosse}}{{Artigau}
  et~al.}{2011}]{artigau2011}
{Artigau}, {\'E}., {Donati}, J.-F.,  \& {Delfosse}, X. 2011, in Astronomical
  Society of the Pacific Conference Series, Vol. 448, 16th Cambridge Workshop
  on Cool Stars, Stellar Systems, and the Sun, ed. C.~{Johns-Krull}, M.~K.
  {Browning}, \& A.~A. {West}, 771

\bibitem[\protect\citeauthoryear{{Bailes} et~al.}{{Bailes}
  et~al.}{2011}]{bailes}
{Bailes}, M., {Bates}, S.~D., {Bhalerao}, V., et~al. 2011, Science, 333, 1717

\bibitem[\protect\citeauthoryear{{Bate}}{{Bate}}{2009}]{Bate_2009}
{Bate}, M.~R. 2009, \mnras, 392, 590

\bibitem[\protect\citeauthoryear{{Bate}, {Bonnell}, \& {Bromm}}{{Bate}
  et~al.}{2003}]{Bate_2003}
{Bate}, M.~R., {Bonnell}, I.~A.,  \& {Bromm}, V. 2003, \mnras, 339, 577

\bibitem[\protect\citeauthoryear{{Boley}}{{Boley}}{2009}]{boley2009}
{Boley}, A.~C. 2009, \apjl, 695, L53

\bibitem[\protect\citeauthoryear{{Bonnefoy} et~al.}{{Bonnefoy}
  et~al.}{2016}]{bonnefoy2016}
{Bonnefoy}, M., {Zurlo}, A., {Baudino}, J.~L., et~al. 2016, \aap, 587, A58

\bibitem[\protect\citeauthoryear{{Boss}}{{Boss}}{1997}]{boss1997}
{Boss}, A.~P. 1997, Science, 276, 1836

\bibitem[\protect\citeauthoryear{{Brandt} et~al.}{{Brandt}
  et~al.}{2014}]{brandt2014}
{Brandt}, T.~D., {McElwain}, M.~W., {Turner}, E.~L., et~al. 2014, \apj, 794,
  159

\bibitem[\protect\citeauthoryear{{Cameron}}{{Cameron}}{1978}]{cameron1978}
{Cameron}, A.~G.~W. 1978, Moon and Planets, 18, 5

\bibitem[\protect\citeauthoryear{{Chabrier}}{{Chabrier}}{2003}]{chabrier2003}
{Chabrier}, G. 2003, \pasp, 115, 763

\bibitem[\protect\citeauthoryear{{Chabrier} \& {Baraffe}}{{Chabrier} \&
  {Baraffe}}{2000}]{chabrier2000}
{Chabrier}, G.,  \& {Baraffe}, I. 2000, \araa, 38, 337

\bibitem[\protect\citeauthoryear{{Chabrier} et~al.}{{Chabrier}
  et~al.}{2014}]{chabrier2014}
{Chabrier}, G., {Johansen}, A., {Janson}, M.,  \& {Rafikov}, R. 2014,
  Protostars and Planets VI, 619

\bibitem[\protect\citeauthoryear{{Claudi} et~al.}{{Claudi}
  et~al.}{2006}]{claudi2006}
{Claudi}, R.~U., {Turatto}, M., {Antichi}, J., et~al. 2006, in \procspie, Vol.
  6269, Society of Photo-Optical Instrumentation Engineers (SPIE) Conference
  Series, 62692Y

\bibitem[\protect\citeauthoryear{{Crepp} et~al.}{{Crepp}
  et~al.}{2012a}]{cre12a}
{Crepp}, J.~R., {Johnson}, J.~A., {Fischer}, D.~A., et~al. 2012a, \apj, 751, 97

\bibitem[\protect\citeauthoryear{{Crepp} et~al.}{{Crepp} et~al.}{2014}]{cre14}
{Crepp}, J.~R., {Johnson}, J.~A., {Howard}, A.~W., et~al. 2014, \apj, 781, 29

\bibitem[\protect\citeauthoryear{{Crepp} et~al.}{{Crepp}
  et~al.}{2012b}]{cre12b}
{Crepp}, J.~R., {Johnson}, J.~A., {Howard}, A.~W., et~al. 2012b, \apj, 761, 39

\bibitem[\protect\citeauthoryear{{Crepp} et~al.}{{Crepp}
  et~al.}{2013a}]{cre13a}
{Crepp}, J.~R., {Johnson}, J.~A., {Howard}, A.~W., et~al. 2013a, \apj, 771, 46

\bibitem[\protect\citeauthoryear{{Crepp} et~al.}{{Crepp}
  et~al.}{2013b}]{cre13b}
{Crepp}, J.~R., {Johnson}, J.~A., {Howard}, A.~W., et~al. 2013b, \apj, 774, 1

\bibitem[\protect\citeauthoryear{{Cumming} et~al.}{{Cumming}
  et~al.}{2008}]{cumming2008}
{Cumming}, A., {Butler}, R.~P., {Marcy}, G.~W., et~al. 2008, \pasp, 120, 531

\bibitem[\protect\citeauthoryear{{Duquennoy} \& {Mayor}}{{Duquennoy} \&
  {Mayor}}{1991}]{duquennoy1991}
{Duquennoy}, A.,  \& {Mayor}, M. 1991, \aap, 248, 485

\bibitem[\protect\citeauthoryear{{Feroz}, {Hobson}, \& {Bridges}}{{Feroz}
  et~al.}{2009}]{MULTINEST}
{Feroz}, F., {Hobson}, M.~P.,  \& {Bridges}, M. 2009, \mnras, 398, 1601

\bibitem[\protect\citeauthoryear{{Fonnesbeck} et~al.}{{Fonnesbeck}
  et~al.}{2015}]{pymc}
{Fonnesbeck}, C., {Patil}, A., {Huard}, D.,  \& {Salvatier}, J. 2015, {PyMC:
  Bayesian Stochastic Modelling in Python}, Astrophysics Source Code Library

\bibitem[\protect\citeauthoryear{{Ford}}{{Ford}}{2005}]{ford2005}
{Ford}, E.~B. 2005, \aj, 129, 1706

\bibitem[\protect\citeauthoryear{{Ford}}{{Ford}}{2006}]{ford2006}
{Ford}, E.~B. 2006, \apj, 642, 505

\bibitem[\protect\citeauthoryear{{Foreman-Mackey} et~al.}{{Foreman-Mackey}
  et~al.}{2013}]{emcee}
{Foreman-Mackey}, D., {Hogg}, D.~W., {Lang}, D.,  \& {Goodman}, J. 2013, \pasp,
  125, 306

\bibitem[\protect\citeauthoryear{{Gelman}, {Roberts}, \& {Gilks}}{{Gelman}
  et~al.}{1996}]{Gelman}
{Gelman}, A., {Roberts}, G.,  \& {Gilks}, W. 1996, Bayesian Statistics, 5, 599

\bibitem[\protect\citeauthoryear{Gelman \& Rubin}{Gelman \&
  Rubin}{1992}]{gelman1992}
Gelman, A.,  \& Rubin, D.~B. 1992, Statist. Sci., 7, 457

\bibitem[\protect\citeauthoryear{{Goldreich} \& {Ward}}{{Goldreich} \&
  {Ward}}{1973}]{goldreich1973}
{Goldreich}, P.,  \& {Ward}, W.~R. 1973, \apj, 183, 1051

\bibitem[\protect\citeauthoryear{Goodman \& Weare}{Goodman \&
  Weare}{2010}]{goodman}
Goodman, J.,  \& Weare, J. 2010, Communications in applied mathematics and
  computational science, 5, 65

\bibitem[\protect\citeauthoryear{{Grossman}, {Hays}, \& {Graboske}}{{Grossman}
  et~al.}{1974}]{grossman1974}
{Grossman}, A.~S., {Hays}, D.,  \& {Graboske}, H.~C., Jr. 1974, \aap, 30, 95

\bibitem[\protect\citeauthoryear{{Hastings}}{{Hastings}}{1970}]{hastings}
{Hastings}, W. 1970, Biometrika, 57, 97

\bibitem[\protect\citeauthoryear{{Hebrard} et~al.}{{Hebrard}
  et~al.}{2016}]{hebrard2016}
{Hebrard}, G., {Arnold}, L., {Forveille}, T., et~al. 2016, ArXiv e-prints,
  1602.04622

\bibitem[\protect\citeauthoryear{Heintz}{Heintz}{1978}]{heintz}
Heintz, W. 1978, Double stars, Geophysics and astrophysics monographs (D.
  Reidel Pub., Co.)

\bibitem[\protect\citeauthoryear{{He{\l}miniak} et~al.}{{He{\l}miniak}
  et~al.}{2016}]{V450And}
{He{\l}miniak}, K.~G., {Kuzuhara}, M., {Mede}, K., et~al. 2016, \apj, 832, 33

\bibitem[\protect\citeauthoryear{{Hogg}, {Myers}, \& {Bovy}}{{Hogg}
  et~al.}{2010}]{hogg2010}
{Hogg}, D.~W., {Myers}, A.~D.,  \& {Bovy}, J. 2010, \apj, 725, 2166

\bibitem[\protect\citeauthoryear{{Ingraham} et~al.}{{Ingraham}
  et~al.}{2014}]{ingraham2014}
{Ingraham}, P., {Marley}, M.~S., {Saumon}, D., et~al. 2014, \apjl, 794, L15

\bibitem[\protect\citeauthoryear{{Juri{\'c}} \& {Tremaine}}{{Juri{\'c}} \&
  {Tremaine}}{2008}]{juric2008}
{Juri{\'c}}, M.,  \& {Tremaine}, S. 2008, \apj, 686, 603

\bibitem[\protect\citeauthoryear{{K{\"a}ufl} et~al.}{{K{\"a}ufl}
  et~al.}{2008}]{CRIRES}
{K{\"a}ufl}, H.~U., {Amico}, P., {Ballester}, P., et~al. 2008, in \procspie,
  Vol. 7014, Ground-based and Airborne Instrumentation for Astronomy II, 70140W

\bibitem[\protect\citeauthoryear{{Kipping}}{{Kipping}}{2013}]{kipping2013}
{Kipping}, D.~M. 2013, \mnras, 434, L51

\bibitem[\protect\citeauthoryear{{Kipping} \& {Sandford}}{{Kipping} \&
  {Sandford}}{2016}]{kipping2016}
{Kipping}, D.~M.,  \& {Sandford}, E. 2016, ArXiv e-prints, 1603.05662

\bibitem[\protect\citeauthoryear{{Kirkpatrick}, {Gelatt}, \&
  {Vecchi}}{{Kirkpatrick} et~al.}{1983}]{SimAnneal}
{Kirkpatrick}, S., {Gelatt}, C.~D.,  \& {Vecchi}, M.~P. 1983, Science, 220, 671

\bibitem[\protect\citeauthoryear{{Kuiper}}{{Kuiper}}{1951}]{kuiper1951}
{Kuiper}, G.~P. 1951, Proceedings of the National Academy of Science, 37, 1

\bibitem[\protect\citeauthoryear{Liu}{Liu}{2008}]{liu}
Liu, J.~S. 2008, {Monte Carlo Strategies in Scientific Computing} (Corrected
  ed.) (Springer)

\bibitem[\protect\citeauthoryear{{Macintosh} et~al.}{{Macintosh}
  et~al.}{2014}]{macintosh2014}
{Macintosh}, B.~A., {Anthony}, A., {Atwood}, J., et~al. 2014, in \procspie,
  Vol. 9148, Adaptive Optics Systems IV, 91480J

\bibitem[\protect\citeauthoryear{{Marley} et~al.}{{Marley}
  et~al.}{2007}]{marley2007}
{Marley}, M.~S., {Fortney}, J.~J., {Hubickyj}, O., {Bodenheimer}, P.,  \&
  {Lissauer}, J.~J. 2007, \apj, 655, 541

\bibitem[\protect\citeauthoryear{{Mede} \& {Brandt}}{{Mede} \&
  {Brandt}}{2014}]{Mede2014}
{Mede}, K.,  \& {Brandt}, T.~D. 2014, in IAU Symposium, Vol. 299, IAU
  Symposium, ed. M.~{Booth}, B.~C. {Matthews}, \& J.~R. {Graham}, 52

\bibitem[\protect\citeauthoryear{{Metchev} \& {Hillenbrand}}{{Metchev} \&
  {Hillenbrand}}{2009}]{metchev2009}
{Metchev}, S.~A.,  \& {Hillenbrand}, L.~A. 2009, \apjs, 181, 62

\bibitem[\protect\citeauthoryear{{Metropolis} et~al.}{{Metropolis}
  et~al.}{1953}]{metropolis}
{Metropolis}, N., {Rosenbluth}, A.~W., {Rosenbluth}, M.~N., {Teller}, A.~H.,
  \& {Teller}, E. 1953, \jcp, 21, 1087

\bibitem[\protect\citeauthoryear{{Mizuno}}{{Mizuno}}{1980}]{mizuno1980}
{Mizuno}, H. 1980, Progress of Theoretical Physics, 64, 544

\bibitem[\protect\citeauthoryear{{Mordasini} et~al.}{{Mordasini}
  et~al.}{2012}]{mordasini2012}
{Mordasini}, C., {Alibert}, Y., {Benz}, W., {Klahr}, H.,  \& {Henning}, T.
  2012, \aap, 541, A97

\bibitem[\protect\citeauthoryear{{Peters} et~al.}{{Peters}
  et~al.}{2012a}]{peters2012}
{Peters}, M.~A., {Groff}, T., {Kasdin}, N.~J., et~al. 2012a, in Society of
  Photo-Optical Instrumentation Engineers (SPIE) Conference Series, Vol. 8446,
  Society of Photo-Optical Instrumentation Engineers (SPIE) Conference Series

\bibitem[\protect\citeauthoryear{{Peters} et~al.}{{Peters}
  et~al.}{2012b}]{CHARIS}
{Peters}, M.~A., {Groff}, T., {Kasdin}, N.~J., et~al. 2012b, in Society of
  Photo-Optical Instrumentation Engineers (SPIE) Conference Series, Vol. 8446,
  Society of Photo-Optical Instrumentation Engineers (SPIE) Conference Series,
  7

\bibitem[\protect\citeauthoryear{{Pollack} et~al.}{{Pollack}
  et~al.}{1996}]{pollack1996}
{Pollack}, J.~B., {Hubickyj}, O., {Bodenheimer}, P., et~al. 1996, Icarus, 124,
  62

\bibitem[\protect\citeauthoryear{{Rodriguez} et~al.}{{Rodriguez}
  et~al.}{2011}]{rodrigues}
{Rodriguez}, D.~R., {Zuckerman}, B., {Melis}, C.,  \& {Song}, I. 2011, \apjl,
  732, L29

\bibitem[\protect\citeauthoryear{{Ryu} et~al.}{{Ryu} et~al.}{2016}]{ryu16}
{Ryu}, T., {Sato}, B., {Kuzuhara}, M., et~al. 2016, ArXiv e-prints, 1603.02017

\bibitem[\protect\citeauthoryear{{Safronov} \& {Zvjagina}}{{Safronov} \&
  {Zvjagina}}{1969}]{safronov1969}
{Safronov}, V.~S.,  \& {Zvjagina}, E.~V. 1969, Icarus, 10, 109

\bibitem[\protect\citeauthoryear{{Shabram} et~al.}{{Shabram}
  et~al.}{2016}]{shabram2016}
{Shabram}, M., {Demory}, B.-O., {Cisewski}, J., {Ford}, E.~B.,  \& {Rogers}, L.
  2016, \apj, 820, 93

\bibitem[\protect\citeauthoryear{{Shen} \& {Turner}}{{Shen} \&
  {Turner}}{2008}]{shen2008}
{Shen}, Y.,  \& {Turner}, E.~L. 2008, \apj, 685, 553

\bibitem[\protect\citeauthoryear{{Tamura} et~al.}{{Tamura} et~al.}{2012}]{IRD}
{Tamura}, M., {Suto}, H., {Nishikawa}, J., et~al. 2012, in Society of
  Photo-Optical Instrumentation Engineers (SPIE) Conference Series, Vol. 8446,
  Society of Photo-Optical Instrumentation Engineers (SPIE) Conference Series,
  1

\bibitem[\protect\citeauthoryear{Tegmark et~al.}{Tegmark
  et~al.}{2004}]{tegmark}
Tegmark, M., Strauss, M.~A., Blanton, M.~R., et~al. 2004, Phys. Rev. D, 69,
  103501

\bibitem[\protect\citeauthoryear{{Thiele}}{{Thiele}}{1883}]{Thiele}
{Thiele}, T. 1883, Astronomische Nachrichten, 104, 225

\bibitem[\protect\citeauthoryear{{Van den Bos}}{{Van den Bos}}{1932}]{Van}
{Van den Bos}, W.~H. 1932, Circular of the Union Observatory Johannesburg, 86,
  261

\bibitem[\protect\citeauthoryear{{van Leeuwen}}{{van
  Leeuwen}}{2007}]{vanLeeuwen_2007}
{van Leeuwen}, F. 2007, \aap, 474, 653

\bibitem[\protect\citeauthoryear{{Vorobyov}}{{Vorobyov}}{2013}]{vorobyov2013}
{Vorobyov}, E.~I. 2013, \aap, 552, A129

\bibitem[\protect\citeauthoryear{{Wang} \& {Ford}}{{Wang} \&
  {Ford}}{2011}]{wang2011}
{Wang}, J.,  \& {Ford}, E.~B. 2011, \mnras, 418, 1822

\bibitem[\protect\citeauthoryear{{Wilson} et~al.}{{Wilson}
  et~al.}{2016}]{wilson2016}
{Wilson}, P.~A., {H{\'e}brard}, G., {Santos}, N.~C., et~al. 2016, ArXiv
  e-prints, 1602.02749

\bibitem[\protect\citeauthoryear{Wolszczan}{Wolszczan}{1994}]{Wolszczan538}
Wolszczan, A. 1994, Science, 264, 538

\bibitem[\protect\citeauthoryear{{Zurlo} et~al.}{{Zurlo}
  et~al.}{2016}]{zurlo2016}
{Zurlo}, A., {Vigan}, A., {Galicher}, R., et~al. 2016, \aap, 587, A57

\end{thebibliography}

\end{document}